\documentclass{pasj00}
\usepackage{comment}
\usepackage{color} 
\draft
\begin{document}
\SetRunningHead{Author(s) in page-head}{Running Head}

\title{High Dispersion Spectroscopy of Solar-type Superflare Stars.  III. Lithium Abundances$^*$}

\author{Satoshi Honda\altaffilmark{1}, Yuta Notsu\altaffilmark{2}, Hiroyuki Maehara\altaffilmark{3}, Shota Notsu\altaffilmark{2}, Takuya Shibayama\altaffilmark{4}, Daisaku Nogami\altaffilmark{2,5}, and Kazunari Shibata\altaffilmark{5}}

\altaffiltext{1}{Nishi-Harima Astronomical Observatory, Center for Astronomy, University of Hyogo, 407-2, Nishigaichi, Sayo-cho, Sayo, Hyogo 679-5313, Japan}
\email{honda@nhao.jp}
\altaffiltext{2}{Department of Astronomy, Kyoto University, Kitashirakawa-Oiwake-cho,Sakyo-ku, Kyoto 606-8502}
\altaffiltext{3}{Okayama Astrophysical Observatory, National Astronomical Observatory of Japan, 3037-5 Honjo,Kamogata, Asakuchi, Okayama 719-0232, Japan}
\altaffiltext{4}{Solar-Terrestrial Environment Laboratory, Nagoya University, Furo-cho, Chikusa-ku, Nagoya, Aichi, 464-8601, Japan}
\altaffiltext{5}{Kwasan Observatory, Kyoto University, Yamashina-ku, Kyoto 607-8471, Japan}


\maketitle

\begin{abstract}
 We report on the abundance analysis of Li in solar-type (G-type main sequence) superflare stars which were found by the analysis of Kepler photometric data. Li is a key element to understand the evolution of the stellar convection zone which reflects the age of solar-type stars. We performed the high dispersion spectroscopy of solar-type superflare stars with  Subaru/HDS, and confirmed that 34 stars show no evidence of binarity in our previous study. In this study, we derived the Li abundances of these 34 objects.
 We investigate correlations of Li abundance with stellar atmospheric parameters, rotational velocity, and superflare activities to understand the nature of superflare stars and the possibility of the nucleosynthesis of Li by superflares.
We confirm the large dispersion in the Li abundance, and the correlation with stellar parameters is not seen.
As compared with the Li abundance in Hyades cluster which is younger than the Sun, it is suggested that half of the observed stars are younger than Hyades cluster.
The measured value of $v \sin i$ (projected rotational velocity) supports those objects are younger than the Sun.
However, there are some objects which show the low Li abundance and slowly rotate on the basis of the estimated $v \sin i$ and $P$ (period of brightness variation).
This result indicates that the superflare stars are not only young stars but also old stars like our Sun.
In our observations, we could not find any evidence of Li productions by superflares.
Further research on Li isotope abundances of superflare stars would clarify the Li production by stellar flares.

\end{abstract}
\footnotetext[*]{Based on data collected at Subaru Telescope, which is operated by the National Astronomical Observatory of Japan.}

\bigskip

\section{Introduction}\label{sec:intro}
Lithium (Li) is easily destroyed in the hotter region (p, $\alpha$ reactions ; $^{7}$Li : $T$ $\geq$ 2.5 $\times$ $10^{6}$ K, $^{6}$Li : $T$ $\geq$ 2.0 $\times$ $10^{6}$ K) of stellar atmosphere.
We can obtain the information about the evolution of convection layer from the behavior of Li abundance.
The accurate spectroscopic determination of the Li abundance in young solar-type stars provide independent and reliable age diagnostics (e.g., \cite{Herbig1965}, \cite{Duncan1981}). 
In addition, accurate measurements of the Li abundance and, in particular, the $^{6}$Li/$^{7}$Li isotopic ratio in stellar atmospheres are of crucial importance for addressing questions about the Big Bang nucleosynthesis (e.g., \cite{Spite1982}, \cite{Boes1985}), the chemical evolution of the Galaxy (e.g., \cite{Ryan2001}), mixing processes in stellar interiors, and the evolution of extra-solar planetary systems (e.g., \cite{Bouvier2008}, \cite{Israelian2009}, \cite{Gonzalez2015}).

Solar flares are the most energetic explosions on the surface of the Sun, and are thought to occur by release of magnetic energy \citep{Shibata2011}.
Flares are also known to occur on various types of stars including solar-type stars.
Among them, young stars, close binary stars, and dMe stars sometimes produce ``superflares", flares whose total energy is 10-$10^{6}$ times larger ($10^{33} - 10^{38}$ erg) than the largest flares on the Sun ($\sim10^{32}$ erg) \citep{Schaefer2000}.
It has been recognized that the superflares will not occur on the slowly-rotating single solar-type stars.
However, \citet{Maehara2012} analyzed the photometric data by the Kepler spacecraft, and discovered 365 superflare events on 148 solar-type stars that have the effective temperature of 5,100K $\leq$  $T_{\rm eff}$ $\leq$ 6,000K, and surface gravity of $\log g$ $\geq$ 4.0.
They also found that the brightness of superflare stars show the quasi-periodic variability with period of from one to a few tens of days.
Those brightness variations can be explained by the rotation of a star with large starspots \citep{YNotsu2013}. This result clearly shows that we can obtain the accurate velocity of stellar rotation. It is important to measure the rotational velocity of stars which is thought to be a key parameter of stellar evolution.
The rotation velocity of a star also reflects the age and activity of a star.
Slowly rotating stars should be quiet and old stars like our Sun.

As a subsequent study, \citet{Shibayama2013} found 1547 superflare events on 279 solar-type stars,
and 44 superflares on 19 slowly rotating ``Sun-like" stars (5,600K $\leq$  $T_{\rm eff}$ $\leq$ 6,000K, $\log g$ $\geq$ 4.0, brightness variation period ($P$) $>$ 10 days).
\citet{Maehara2015} also found 187 superflares on 23 solar-type stars using Kepler data with 1 min sampling.
With these data, we studied the statistical properties of the occurrence rate of superflares, and found that the occurrence rate ($dN/dE$) of superflares versus flare energy ($E$) shows a power-law distribution with $dN/dE \propto E^{-\alpha}$ where $\alpha \sim 2$ \citep{Maehara2012,Shibayama2013,Maehara2015}.
It is interesting that this distribution is roughly similar to that for solar flares.
These results indicate that it is quite likely that superflares would occur on the Sun.
In addition to this research, \citet{Shibata2013} show the possibility of occurrence of superflares on the present Sun from the theoretical point of view.
\citet{SNotsu2013} show the chromospheric activity of the solar-type superflare star (KIC 6934317) by high dispersion spectroscopy.
They found high chromospheric activity from the core depth of Ca II and H$\alpha$ lines of KIC 6934317.
These results support the idea of existence of large starspots.
They also found the low Li abundance in that star.
\citet{Nogami2014} obtained spectroscopic data of the slowly-rotating superflare stars KIC 9766237 and KIC 9944137, and found that these superflare stars have very Sun-like atmospheric parameters including the Li abundance of the Sun.
\citet{Wichmann2014} carried out high dispersion spectroscopy for superflare stars from the list of \citet{Maehara2012}.
They found that several of those stars are very young and active, but some other stars do not show the cause of occurrence of superflares.

Notsu et al. (2015a,b; hereafter Paper I, II) observed a large sample of ``solar-type" superflare stars.
They picked up single stars and confirmed the atmospheric parameters of ``solar-type" superflare stars are similar to the solar ones by high dispersion spectroscopy. 
In particular, they found that 9 stars are in the range of ``Sun-like" stars.
This result supports the hypothesis that the Sun might cause a superflare.
In paper II, they showed the correlation between the intensity of Ca II infrared triplet lines, which are good indicators of the stellar chromospheric activities, and amplitude of brightness variation among their observed superflare stars.
These results clearly show that all the targets have large starspots and high chromospheric activity compared with the Sun.
From those observational results, we can say that the brightness variation of superflare stars is due to the rotation with large spots.
The problem which we have to consider next is whether the present Sun causes the superflare or not.

Several studies have been done on the behavior of Li abundances in solar-type stars so far.
Takeda et al. (2007, 2010) observed the large sample of ``solar-analog" stars (selected with the photometric criteria of $0.62 \lesssim$ $B-V$ $\lesssim 0.67$ and $4.5 \lesssim M_v \lesssim 5.1$) and investigated the correlations of stellar parameters with Li abundance.
They found clear correlations among Li abundance, stellar activity and rotation.
\citet{Mishenina2012} also investigated the correlations among Li abundance, stellar parameters and activity in the F,G,K dwarfs.
They show the Li-activity correlation is evident only in a restricted temperature range (5,200 $<$  $T_{\rm eff}$ $<$ 5,700K), and the spread of Li abundance seems to be present in a group of low chromospheric activity stars.

The possibilities of Li production in stellar flares have been suggested by some researches.
\citet{Canal1974} shows the possibility of nucleosynthesis of Li in low-energy flares in T Tauri-like stars.
\citet{KarpenWorden1979} estimated the amounts of Li produced by stellar-flare processes on UV Ceti stars.
They conclude that less than 10 \%
 of the $^{7}$Li observed in the interstellar medium are caused by stellar flares.
\citet{delareza1981} determined the Li abundance in flare stars.
They did not find a general relation between Li abundance and chromospheric activity.
On the other hand, \citet{Tatischeff2007} show that the $^{6}$Li could be produced by the reaction $^{4}$He($^{3}$He, $p$)$^{6}$Li in situ by solar-like flares.
\citet{MontesRamsey1998} reported the Li line enhancement during a long-duration flare in active binary.
They suggest this Li enhancement is caused by spallation reactions during the flare.
However, there is no clear evidence of productions of Li by stellar flares.
In addition, \citet{Giampapa1984} demonstrated that the spots and plages could alter the observed Li line on the basis of solar observations.
Although many attempts have been made to study the possibilities of Li production in stellar flares, it is still controversial.

Here we show the lithium abundance in superflare stars and discuss it's relation with stellar activity, rotation velocity, and age of stars.
In paper I and II, we determined the atmospheric parameters of superflare stars with high dispersion spectroscopy and discuss the spots and activity for understanding the nature of superflare stars.
This is the third paper of our spectroscopic studies of superflare stars.
In Sect. 2 and 3, we describe the observations and analysis of Li abundance in superflare stars, and in Sect. 4 we show the behaviors of Li abundance as a function of stellar parameters, discuss the reasons of Li distribution and the possibility of Li production by superflares.
In Sect. 5 we summarize the paper.

\section{Observations}\label{sec:tarobs}

High dispersion spectroscopy was carried out for 50 solar-type superflare stars (Notsu et al. 2013a, 2015a), which were found by \citet{Maehara2012} and \citet{Shibayama2013}, on 2011 Aug. 3 (S11B-137S), 2012 Aug. 6-8, Sep. 22-25 (S12B-111N) and 2013 June 23-24 (S13A-045N) using Subaru High Dispersion Spectrograph (HDS : \cite{Noguchi2002}).
We applied the image slicer ($\sharp$2) in the observations of S13A \citep{Tajitsu2012}.
In addition, we have observed 10 bright solar-type stars and the Moon for comparison.
Among them, 8 stars are reported as ``solar-twin" stars by the previous studies (King et al. 2005, Takeda \& Tajitsu 2009, Datson et al. 2012).
The details of observations and target stars are shown in Paper I.

The spectrum covers 6100-8820 \AA~ with a resolving power ($R = \lambda / \Delta \lambda$) of 97,000 for S11B, 51,000 for S12B and 80,000 for S13A by 2 $\times$ 2 on-chip binning.
This spectral range includes lines of Li I (6708 {\AA}), H$\alpha$ (6563 {\AA}), and Ca II triplet (8498, 8542, 8662 {\AA}).
The reduction was carried out in a standard manner using the IRAF echelle package\footnote{IRAF is distributed by the National Optical Astronomy Observatories, which is operated by the Association of Universities for Research in Astronomy, Inc. under cooperative agreement with the National Science Foundation.}.
We have excluded the binary stars in the following analysis.
The 16 of 50 targets were recognized to be binary systems which show double (or multiple) lines, variations of radial velocity or having visual companion stars (cf., Paper I).

\section{Analysis}\label{sec:tarobs}

We derived the Li abundances and $v \sin i$ (projected rotational velocity) from those spectra using the atmospheric parameters determined by Paper I.
The method of estimations of $v \sin i$ is described in \citet{SNotsu2013} and Paper I.
The objects are distributed in the range of $T_{\rm eff}$ from 5,000K to 6,300K and $\log g$ from 3.5 to 4.9.
We note that there are large differences between atmospheric parameters taken from Kepler Input Catalog (KIC : \cite{Brown2011}) and our obtained values (Paper I).
The estimated values of atmospheric parameters by spectroscopy are more reliable than that from photometric data such as KIC values (cf., Paper I).
The adopted values in this analysis are shown in Table 1.

For the abundance analysis, we used the analysis program SPTOOL\footnote{http://optik2.mtk.nao.ac.jp/$\sim$takeda/sptool/} which was developed by Y. Takeda (private communication), based on Kurucz's ATLAS9/WIDTH9 \citep{kurucz93}.
We assumed local thermodynamic equilibrium (LTE) and derived abundances using the synthesis spectrum with interpolated model atmospheres taken from \citet{kurucz93}.
The line data around Li I 6708 {\AA} region are taken from the list of \citet{TakedaKawanomoto2005}, which are based on the list of \citet{Smith1998} and \citet{Lambert1993}.
We assumed the contributions of $^{6}$Li are negligible.
Figure 1 shows the spectra of all sample stars in the portion of the Li I 6708 line region.
We used the value of solar abundances obtained by \citet{Asplund2009}.
The upper limit of Li abundance is estimated by the method of \citet{TakedaKawanomoto2005}.
They used the averaged FWHM (full width at half maximum) of weak Fe lines and signal to noise ratios (S/N) for deriving the upper limit value.
Adopted atmospheric parameters and derived Li abundances are shown in Table~1.

We assume that there are two main sources of error.
One is a systematic error by atmospheric parameters, and another is a random error depending on S/N.
Systematic errors are caused by the choice of the model atmosphere.
The typical estimated errors for $T_{\rm eff}$, $\log g$, $v_{\rm turb}$, and [Fe/H] are 50 K, 0.2 dex, 0.2 km s$^{-1}$, and 0.1 dex on the basis of Paper I.
Changes in the abundance of Li caused by errors of the above parameters are 0.05, 0.01, 0.06, and 0.00 dex for the typical atmospheric parameter.
The total systematic error is 0.08 dex, which is estimated from the root sum square of the above 4 systematic error values.

The uncertainties arising from profile fitting error depend on the S/N and equivalent width of Li I line.
We estimate the errors of measurement of equivalent width by using the formula derived by \citet{Cayrel1988}, which is taking in the S/N, resolution, and FWHM of line.
Then, we estimate the severity of the value of $A$(Li) from that estimation.
In most cases, the changes of $A$(Li) are smaller than 0.05 dex.
However, a few objects show larger than 0.1 dex.
The largest value is 0.3 dex for KIC~7354508.
It is necessary to keep in mind that the low $A$(Li) stars have larger error than that of high $A$(Li) stars.

In the case of typical atmospheric parameter and Li abundance ($T_{\rm eff}$ = 5,500K, $\log g$ = 4.4, $A$(Li) = 2.1) in our sample, the effect of non-LTE is about 0.06 dex by using the grid of corrections \citep{Carlsson1994}.
However, it must be noted that the correction of --0.3 dex is the most effective case for KIC~11610797, which has the highest $T_{\rm eff}$ and Li abundance in our sample.

We then assume the typical error of Li abundance is about 0.15 dex in those objects from what has been discussed above.

\section{Results and Discussion}\label{sec:discussion}
\subsection{Behaviors of Li abundance with effective temperature ($T_{\rm eff}$)}\label{subsec:dis-binarity}

Figure 2 shows the behavior of the Li abundances as a function of $T_{\rm eff}$ of our target stars with F,G,K type stars \citep{TakedaKawanomoto2005}.
The Li depletion is remarkably seen in the stars whose temperature is lower than the Sun ($T_{\rm eff}$ $\lesssim$ 5,500K).
On the other hand, the main sequence stars having a temperature higher than 6,000K shows no Li depletion.
This is because the evolution of the convection zone in the stellar atmosphere make transport the Li to deeper zone, and destruction of Li in the lower $T_{\rm eff}$ stars.
The depletion of Li in the stellar surface by convective mixing increases with a lapse of time.
Hence, we can consider that high Li stars are young stars.

However, a large diversity (by more than 2 dex) of the Li abundance is seen in stars with the solar temperature, despite the similarity of stellar parameters (cf., \cite{TakedaKawanomoto2005}).
Especially, the solar Li abundance is quite low.
The standard stellar model cannot explain this diversity.
The behavior of the Li abundance in solar-analog stars is still unclear.

The Li abundances of superflare stars do not show clear relations with $T_{\rm eff}$.
The temperature distribution of our target stars is slightly lower than that of the Sun.
We have selected the target of superflare stars having the parameters similar to the Sun based on the temperature of KIC.
The estimated temperature by KIC is systematically about 200K lower than other studies (e.g., \cite{Pinsonneault2012}, Paper I).

We can find the trend of low Li for rotation period ($P$) $\geq$ 10 days and high Li for $P <$ 10 days.
This trend is the same for the superflare stars having the parameters similar to the Sun (Figure 2b).
The stellar rotation also might affect the Li depletion in superflare stars.

We can consider that three stars (KIC 11610797, KIC 9652680, and KIC 8429280) which show an especially high value of Li are very young stars.
It is also supported that these stars are young, because they show large projected rotational velocity ($v \sin i$) and small rotation period ($P$).
In addition to these three stars, about half of the objects show high Li compared with the stars in the Hyades cluster (Figure 3).
The estimated age of Hyades cluster is 6.25 $\times$ 10$^{8}$ yr (e.g., \cite{Perryman1998}), which is one of the old open clusters but younger than the solar age.
It is acceptable to suppose that the young stars show high activities and superflares.

However, more than ten stars with superflares do not show high values of Li (including upper limits) compared with the Hyades cluster.
In particular, some objects show quite low values of Li like the Sun.
Among them, stellar parameters of KIC 9766237 and KIC 9944137 are also very similar to the Sun \citep{Nogami2014}.

\subsection{Rotational velocity and Li abundance}

\citet{Maehara2012} found the frequency of superflares increases with increasing the stellar rotation velocity.
In general, the stellar rotation has a correlation with the age and activity of the star.
The young stars have rapid rotation, high activity and high Li abundance, but they have the decrease in the rotation, the activity and Li abundance with the age \citep{Skumanich1972}.

\citet{Takeda2010} proposed that the stellar rotation ($v \sin i$) may be the most important parameter in determining the surface Li content in the solar-analog stars.
Figure 4 shows the correlation of Li abundance with $v \sin i$ of superflare stars and ordinary solar-analog stars \citep{Takeda2010}.
The $v \sin i$ of solar-analog stars distributed mainly between from 2 to 5 km s$^{-1}$, and which show a clear correlation with Li abundance.

Superflare stars show a wider range of $v \sin i$ than the solar-analog stars.
We must note that this is not the typical distribution of superflare stars due to the selection bias (Paper I).
However, we can find the trend of low Li and large $v \sin i$ in superflare stars compared with solar-analog stars.
Some of them have a lower temperature ($T_{\rm eff}$ $\leq$ 5,500K) and gravity ($\log g$ $<$ 4) than those of the Sun.
Those stars tend to show lower Li abundances since they have the convection layer deeper than the typical solar-analog stars (5,500 K $<$ $T_{\rm eff}$ $<$ 6,000K, $\log g$ $\geq$ 4).
It should also be added that those stars should be rapid rotation stars ($v \sin i$ $>$ 5 km/s and $P$ $<$ 10 days).

Exception is KIC 11764567, which shows large $v \sin i$ and low Li, but it has Sun-like atmospheric parameters.
The rotational velocity of this star estimated from $P$ is inconsistent with $v \sin i$ (cf., Paper II).
This star might belong to a binary system.
In the case of a binary system, observed superflares do not always reflect the primary solar-type star's phenomenon.

The Li abundances of KIC 4831454 and KIC 7354508 are remarkably deviated from the solar-analog stars.
KIC 4831454 has a small inclination angle (Paper II), which should be rapid rotation by the estimation from $P$.
On the other hand, KIC 7354508 shows lower Li abundance than the solar one in spite of slightly larger $v \sin i$ than the solar one.
We also found that the star shows a large value of radial velocity ( --112.8 km s$^{-1}$ ; Paper I ).
It is quite likely that this star is also belongs to a binary system.

We can find that there are some superflare stars with low Li abundance ($A$(Li) $<$ 1.5) and small $v \sin i$ ( $\lesssim$ ~2 km s$^{-1}$)
like the Sun.
Among them, KIC 1197517, KIC 8359398, KIC 11303472, and KIC 6504503 are not Sun-like stars.
KIC 1197517, KIC 8359398, and KIC 11303472 have low temperature ($T_{\rm eff}$ $<$ 5,300K) and KIC 6504503 has low gravity ($\log g$ $=$ 3.6).
However, there are some Sun-like superflare stars which show slow-rotation and low Li abundance like the Sun.

In Paper I, we tried to estimate the age of superflare stars from isochrone.
It is difficult to distinguish whether slightly low $\log g$ stars are post-MS or pre-MS only from the isochrone.
Li abundance gives important clues about the age of the star.
The Li abundances in many superflare stars have higher values than that of the Sun. 
Especially, those stars with a very high Li abundance are suggested to have a rapid rotation velocity from large $v \sin i$ and brightness variations having a short period. 
The stellar rotation is also an index of the age.
It seems reasonable to assume that the age of those stars are young. 
However, some superflare stars show small $v \sin i$ (slow rotation) and low Li abundance. 
This result may indicate that those stars are not young.
Superflare stars are not necessarily young on the basis of our spectroscopic observations.

\subsection{Ca II ($r_{0}$(8542)) indexes and Li abundances.}

The $r_{0}$(8542) index is residual core flux normalized by the continuum level at the line core of the Ca II (8542 \AA).
The $r_{0}$(8542) index reflects the activity of stellar chromosphere (e.g., \cite{Linsky1979}) and well correlates with the intensity of the stellar mean magnetic field (cf., Paper II).
\citet{Takeda2010} shows the positive correlation among Li abundances, $r_{0}$(8542) and $v \sin i$ for solar-analog stars.
They conclude the depletion of surface Li in solar-type stars operates more efficiently as stellar rotation decelerates.

In Paper II, the measured $r_{0}$(8542) index shows that the superflare stars have higher chromospheric activity compared with the Sun.
In addition to this, we found the correlation between the amplitude of the brightness variation and the $r_{0}$(8542).
This correlation can be explained by the difference of the spot size of superflare stars.

Figure 5 shows the Li abundances as a function of $r_{0}$(8542) index for solar-analog stars \citep{Takeda2010}.
We could not find the positive correlation between Li abundance and $r_{0}$(8542) of superflare stars compared with solar-analog stars.
\citet{Duncan1981} investigated the Li abundances and core flux of Ca II K line in solar-type stars.
They found there are a number of high Li abundances and very little chromospheric emission flux, and the converse is rare.
However, there are some low Li and high $r_{0}$(8542) stars, which are not rare in superflare stars.

These results indicate that some superflare stars have large active region but low Li abundance.
Our results are not consistent with the Skumanich's law \citep{Skumanich1972}.
The Skumanich's law cannot apply to these superflare stars.
We cloud not determine whether those stars are young or old from Ca II flux.

The large spots are constituted to make enhancement of Li abundance than the normal region.
\citet{Pallavicini1987} suggested that the strong Li line could be due to large cool spots in RS CVn stars.
Superflare stars should have large spots on the stellar surface estimated from the amplitude of brightness variations.
However, we can not find the strong Li line in such superflare stars with large spots.
That effect may not be seen on about region of 10\% 
(e.g., Notsu et al. 2013b) in the stellar surface.

\subsection{The possibility of Nucleosynthesis of Li in a Superflare.}

\citet{Tatischeff2007} showed the possibilities of Li ($^{6}$Li) production by stellar flares.
However, there is no conclusive evidence that Li is  produced in situ by stellar flares.
Superflare stars are good objects to investigate whether flares can make Li or not.
If Li production occurs during superflares, we could see an increase in the Li abundance of the superflare stars with frequent large flares.
Contrary to such hypothesis, the results indicated that the Li abundance decreases as the number of superflares increases.
Figure 6 shows the Li abundance as a function of the number of superflares and frequency of superflares in our target stars.
We must note that the numbers of superflares should be a lower limit.
It is because the detection of relatively small superflares is somewhat affected by stellar brightness variations \citep{Shibayama2013}.
The frequency of superflares is obtained by dividing the number of superflares by the total observation time.
KIC 8547383, KIC 7420545, KIC 6934317, and KIC 11764567 had superflares many times ($> 30$) in spite of their not high value of Li abundances ($A$(Li) $<$ 2).

The maximum energy of the superflare also does not show any correlation with the Li abundance.
For example, KIC 10471412 shows the largest maximum energy of the superflare ($5.2 \times 10^{35}$ erg : Paper I), but Li abundance is not high ($A$(Li) $<1.5$).

In our observations, we could not find any evidence of Li production by superflares.
However, our observations could not show the small contributions of $^{6}$Li.
Further research on Li isotope abundances of superflare stars would clarify the Li production by stellar flares.
In order to know the reason for the high Li abundance of young stars, it is important to investigate the ratio of $^{6}$Li to $^{7}$Li in the solar-analog superflare stars. 

\section{Summary}

We have estimated the Li abundance of superflare stars and investigate the correlations of Li abundance with stellar parameters, and the possibility of Li productions in stellar flares.
Our spectroscopic observations show the slightly young solar-type stars tend to produce superflares, but old superflare stars exist.
There is a possibility that superflares would be generated on our Sun.
We could not find any evidence of nucleosynthesis of Li in stellar flares from our observations.

\bigskip
\bigskip
We would like to thank Dr. Yoichi Takeda for useful comments and providing the analysis tools.
This study is based on observational data collected with Subaru Telescope, which is operated by the National Astronomical Observatory of Japan.
We are grateful to Dr.~Akito Tajitsu and other staffs of the Subaru Telescope for making large contributions in carrying out our observation.
We also thank an anonymous referee for helpful comments.
Funding for this mission is provided by the NASA Science Mission Directorate.
The Kepler data presented in this paper were obtained from the Multimission Archive at STScI. 
This work was supported by the Grant-in-Aids from the Ministry of Education, 
Culture, Sports, Science and Technology of Japan (No. 25287039, 26400231, and 26800096).

\clearpage

\begin{figure}[htbp]
 \begin{center}
  \FigureFile(75mm,75mm){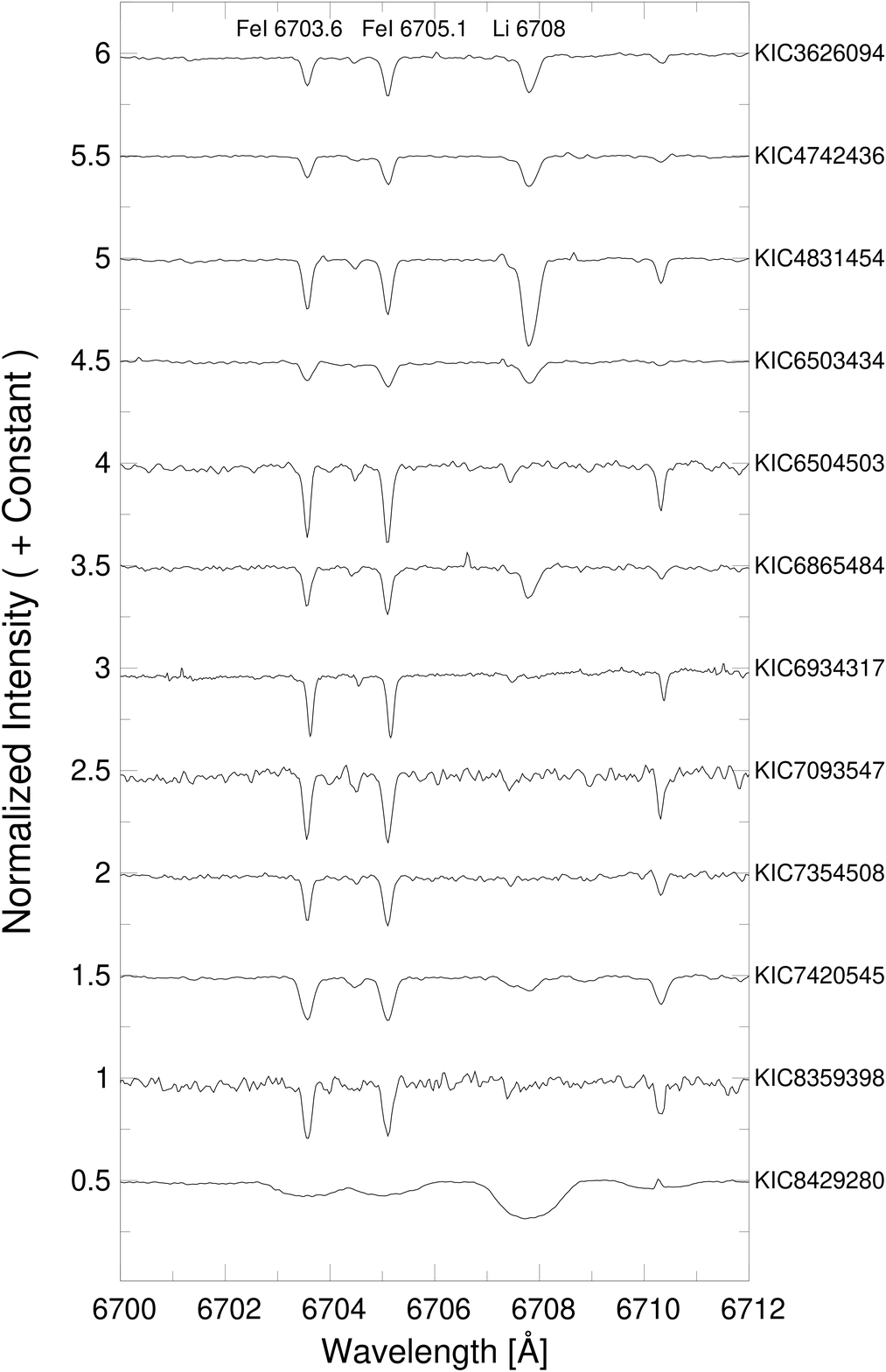}
  \FigureFile(75mm,75mm){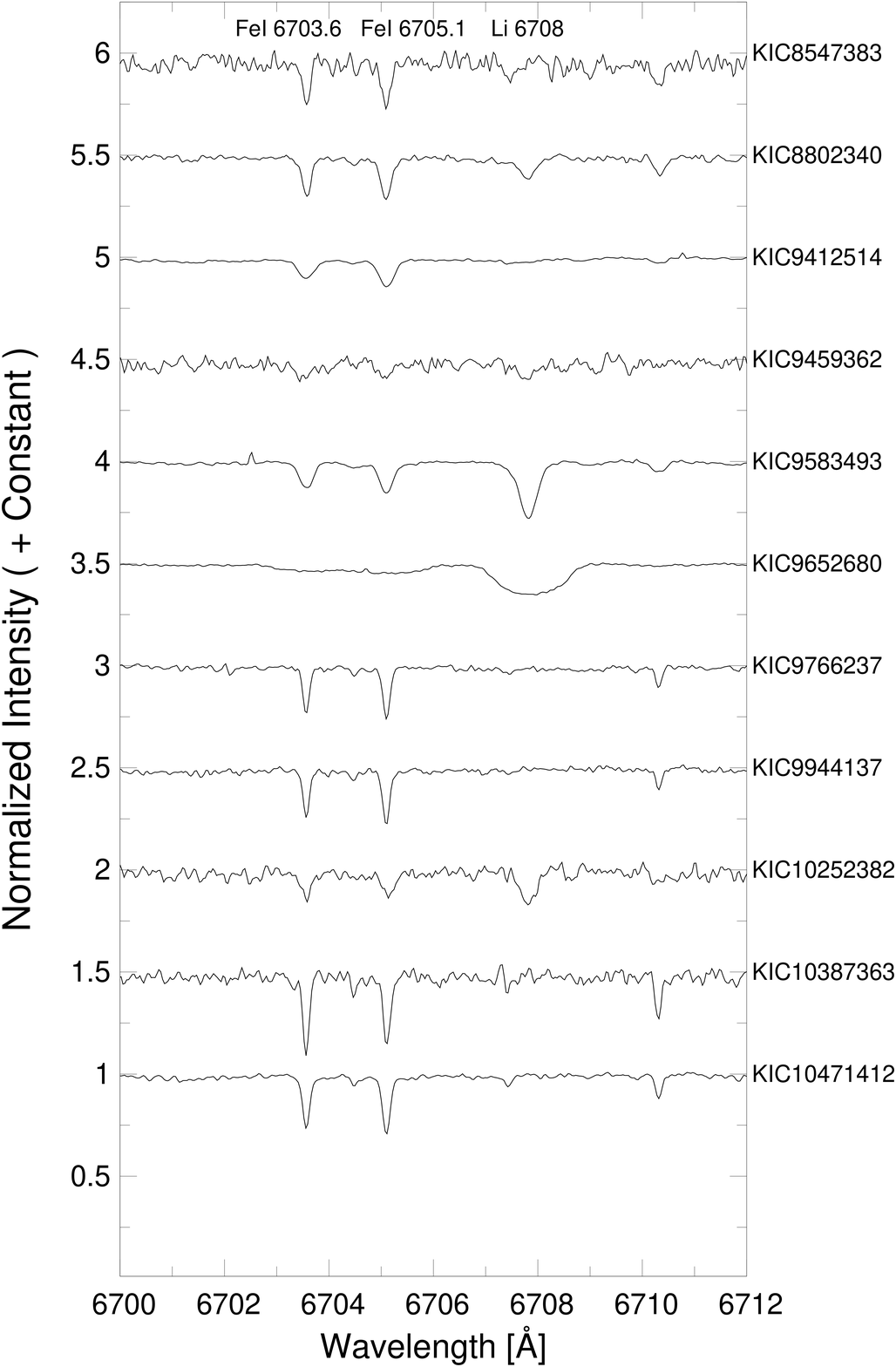}
  \FigureFile(75mm,75mm){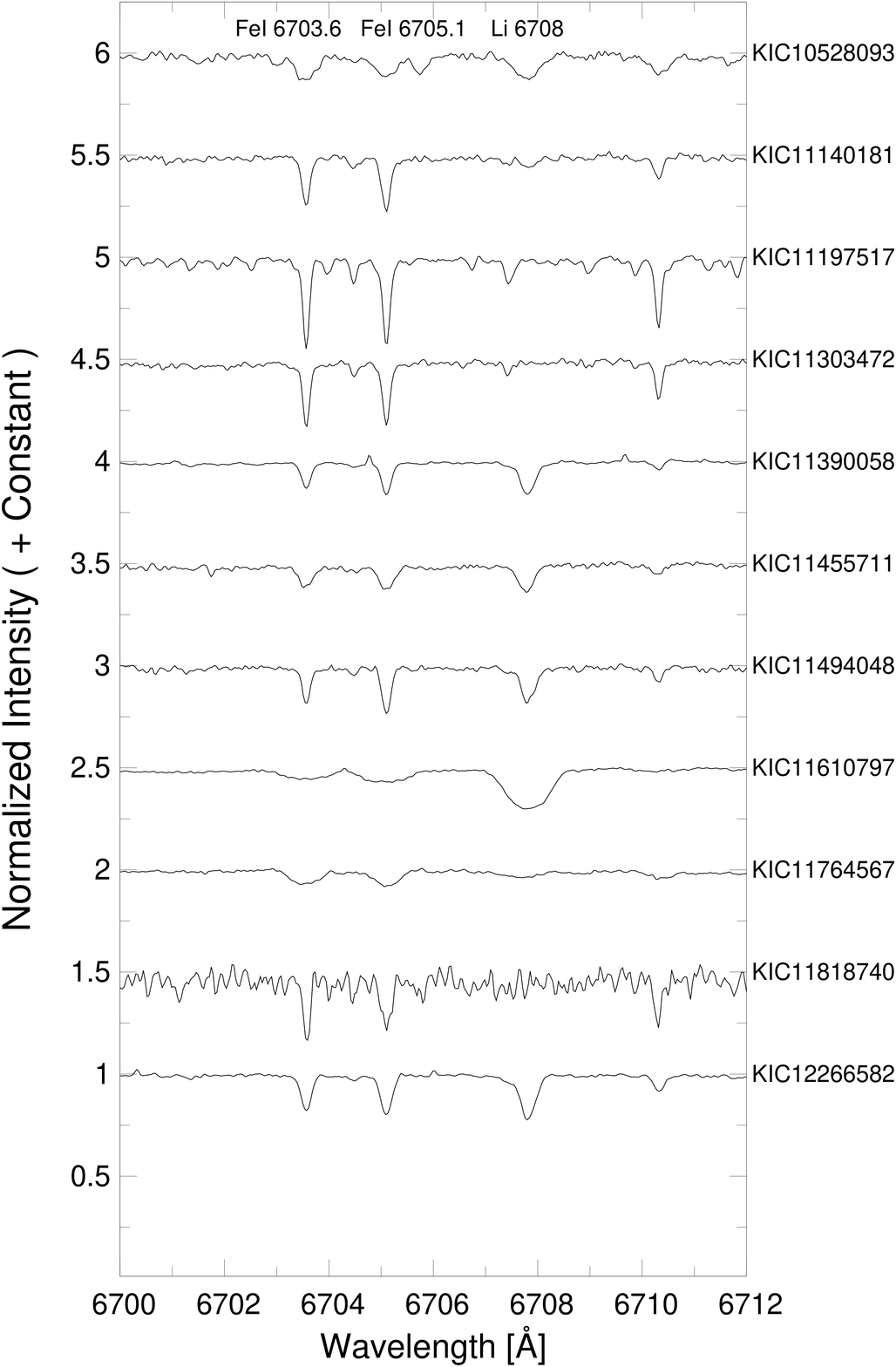}
  \FigureFile(75mm,75mm){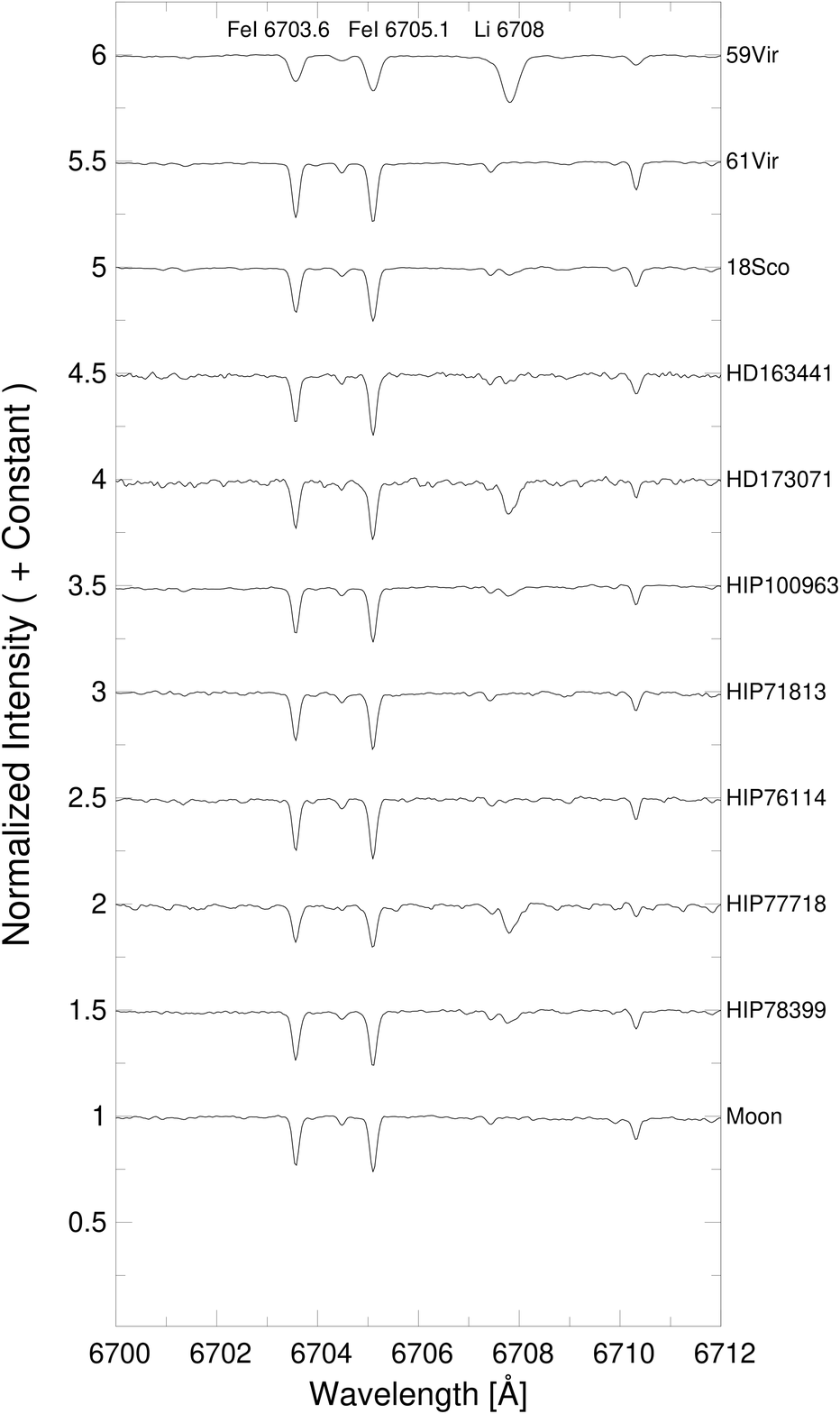}
 \end{center}
\caption{Spectra around Li I 6708 line of the 34 superflare stars that show no evidence of binarity, 
10 comparison stars, and the Moon.
The wavelength scale is adjusted to the laboratory frame. 
Co-added spectra are used here in case that the star was observed multiple times.
}\label{fig:sp-Li}
\end{figure}

\begin{figure}[htbp]
 \begin{center}
  \FigureFile(120mm,120mm){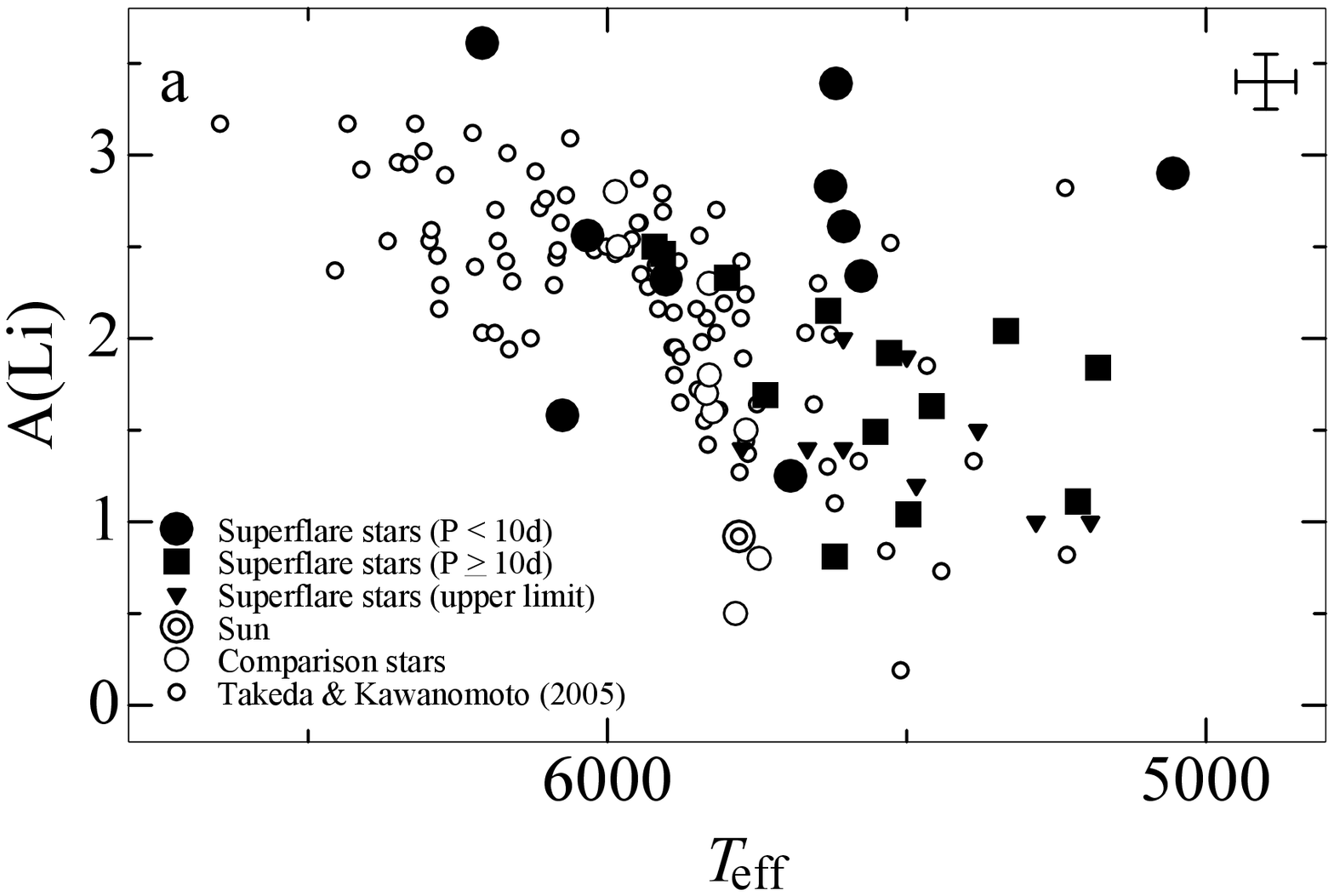} 
  \FigureFile(120mm,120mm){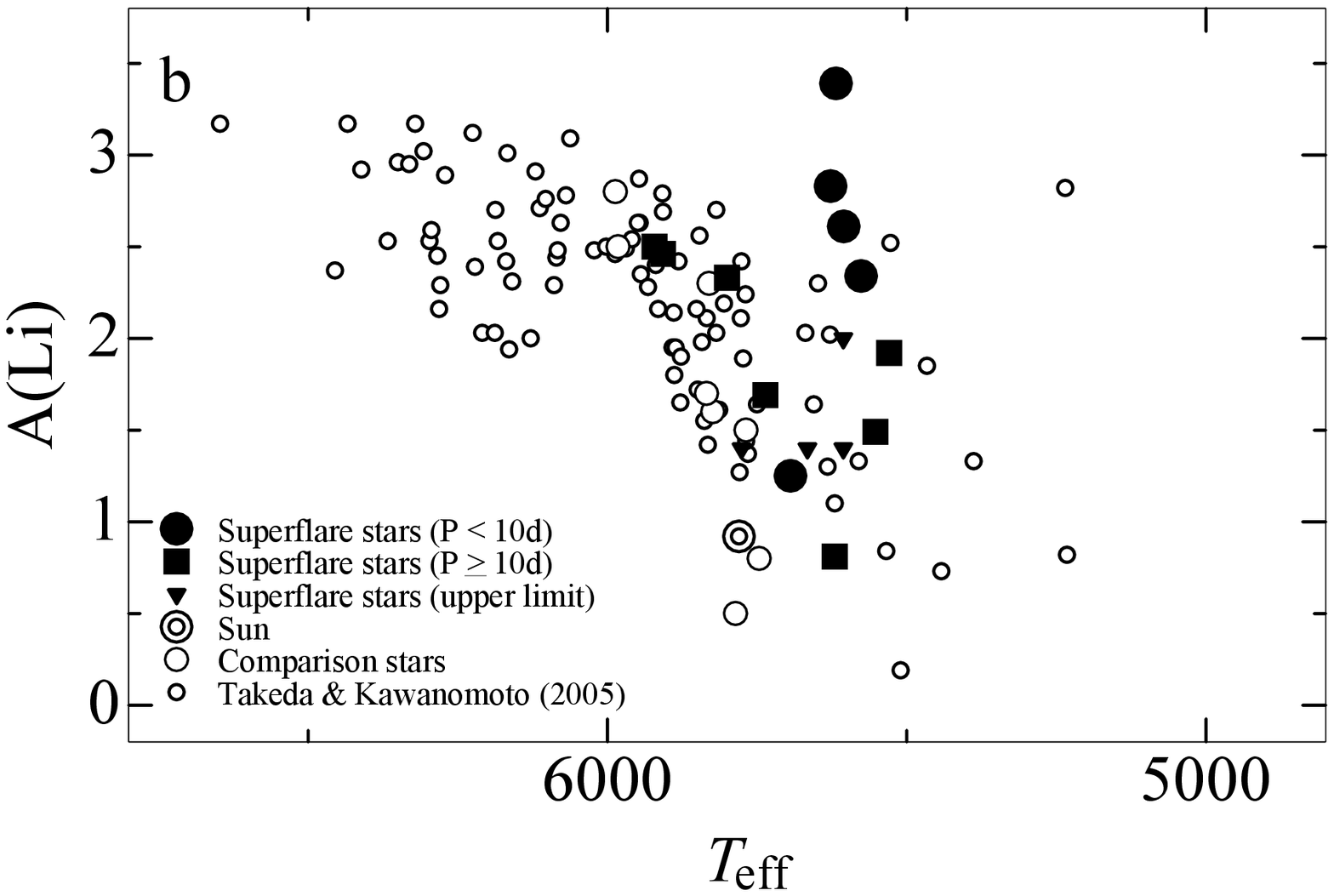} 
 \end{center}
\caption{Lithium abundances ($A$(Li)) vs. effective temperatures ($T_{\rm eff}$). {\bf a} : Filled circles and squares indicate the Li abundance of superflare stars with $P < $ 10 and $P$ $\geq$ 10 days respectively, and triangles also indicate the upper limit of Li  abundance of superflare stars. Open small circles indicate the ordinary F, G, K-type main sequence stars (Takeda \& Kawanomoto 2005), and open large circles indicate comparison stars in this observation. {\bf b} :All symbols are the same as {\bf a}, but the data points of superflare stars are limited to solar-analog (5,500K $<$ $T_{\rm eff}$ $<$ 6,000K, $\log g$ $\geq$ 4) superflare stars.}\label{fig:exbi}
\end{figure}

\begin{figure}[htbp]
 \begin{center}
  \FigureFile(120mm,120mm){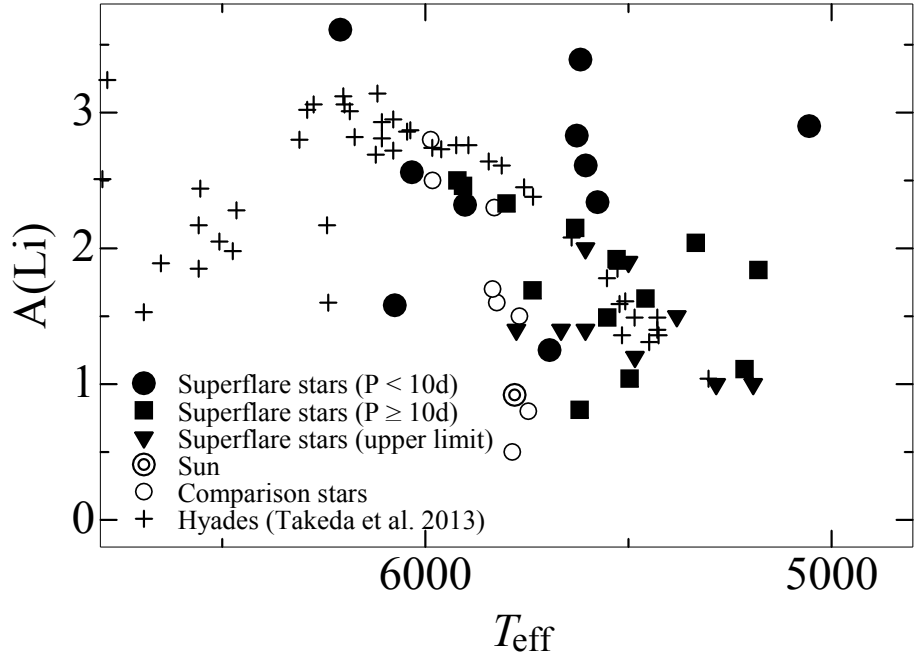} 
 \end{center}
\caption{$A$(Li) vs. $T_{\rm eff}$ with stars in the Hyades cluster. Upper panel : The symbols of filled circles and squares are the same as in Figure 1, but crosses indicate the stars of the Hyades cluster (Takeda et al. 2013). The age of Hyades is 6.25 $\times$ 10$^{8}$ yr (e.g., Perryman et al. 1998).
}\label{fig:exbi}
\end{figure}

\begin{figure}[htbp]
 \begin{center}
  \FigureFile(120mm,120mm){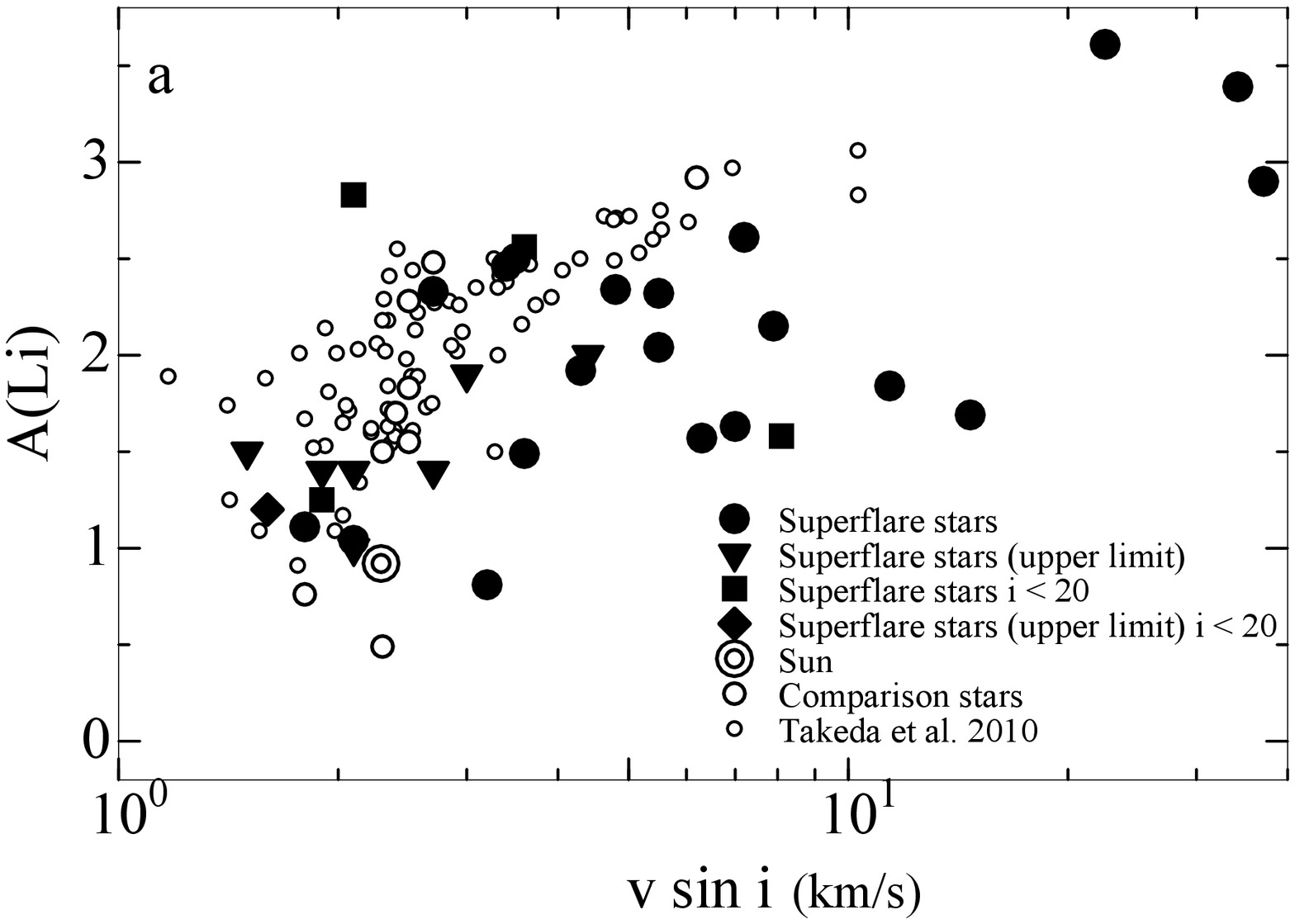} 
  \FigureFile(120mm,120mm){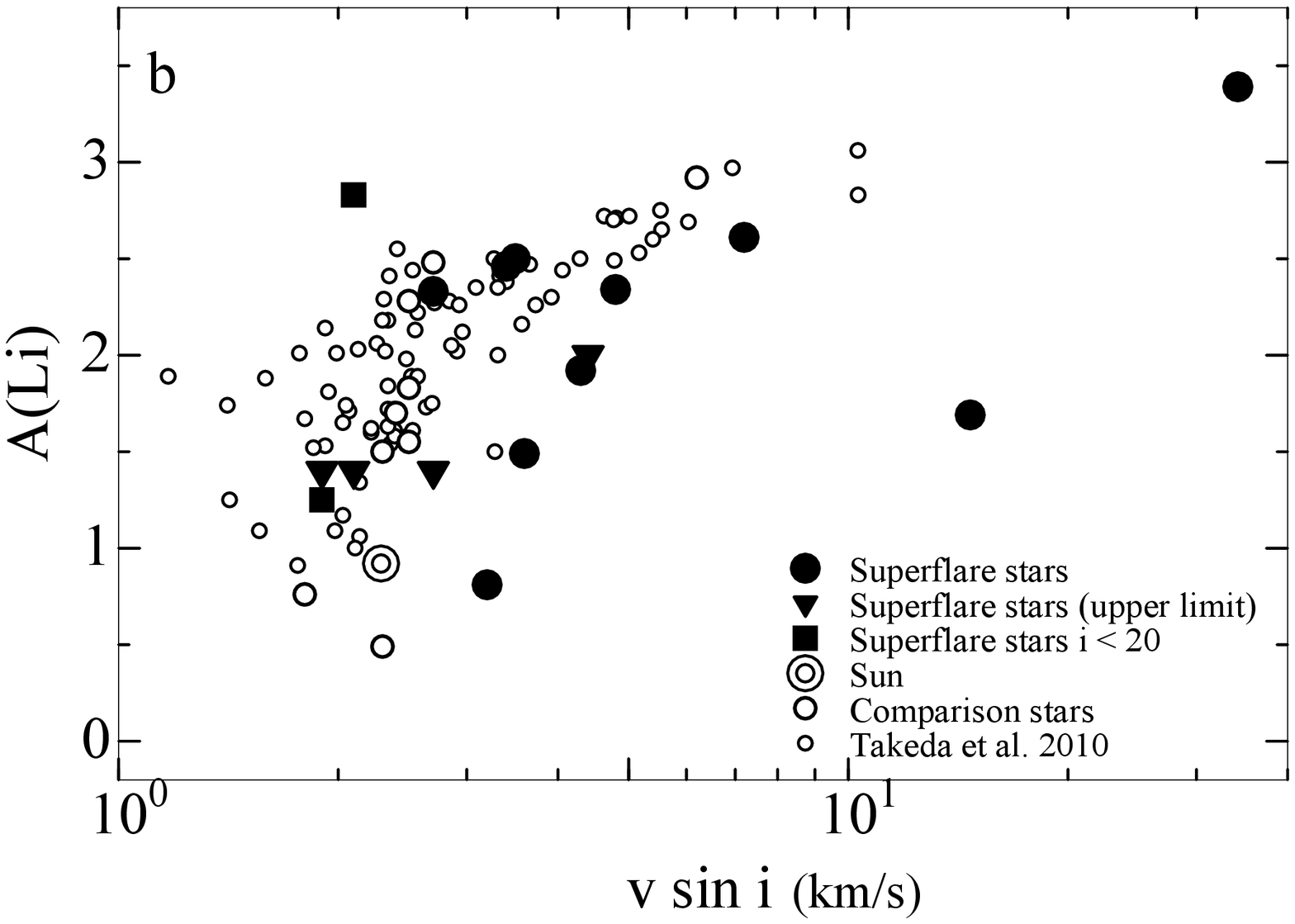} 
 \end{center}
\caption{$A$(Li) vs. $v \sin i$. {\bf a} : The symbols of filled circles, triangles, squares, and diamond indicate the measurements of this work.
The plot of small open circles are taken from Takeda et al. (2010), which are the solar-analog stars.
The small inclination angle stars ($i < 20$ ; Paper II) are shown as squares and diamond. {\bf b} : All symbols are the same as {\bf a}, but the data points of superflare stars are limited to solar-analog (5,500K $<$ $T_{\rm eff}$ $<$ 6,000K, $\log g$ $\geq$ 4) superflare stars.}\label{fig:exbi}
\end{figure}

\begin{figure}[htbp]
 \begin{center}
  \FigureFile(120mm,120mm){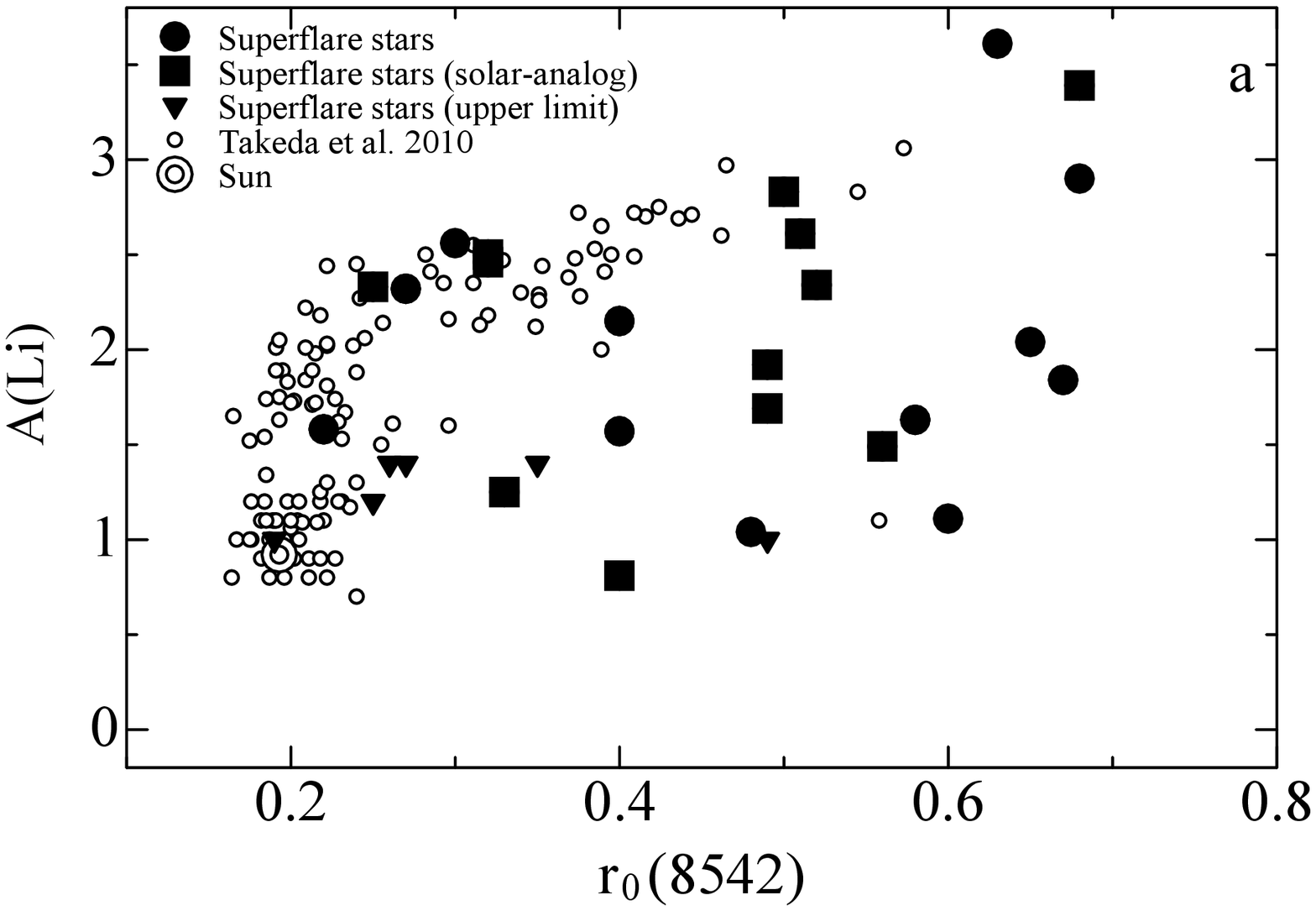} 
  \FigureFile(120mm,120mm){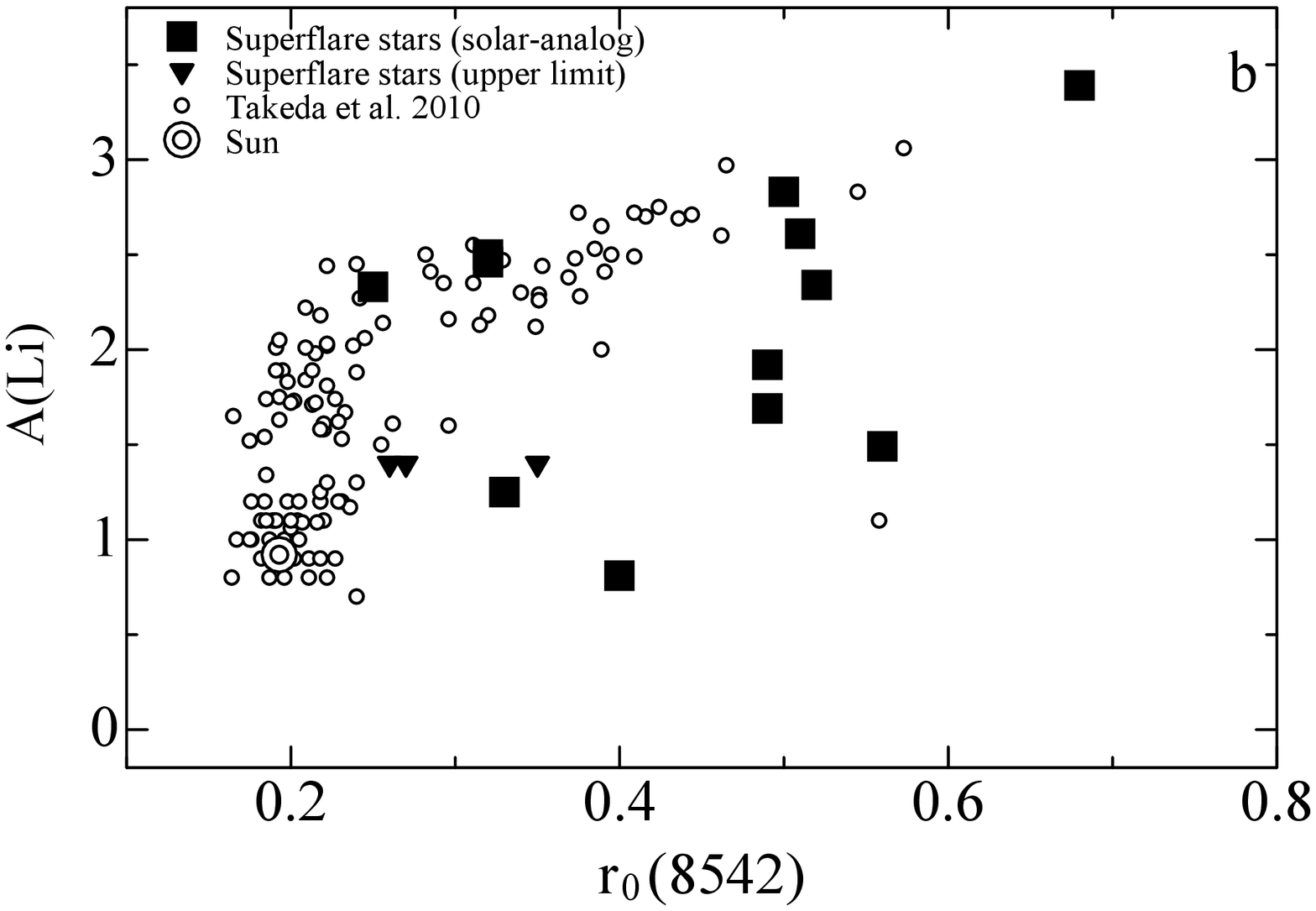} 
 \end{center}
\caption{$A$(Li) vs. $r_{0}$(8542). {\bf a} : The symbols of filled circles, squares, and triangles indicate the measurements of this work. The plot of small open circles are taken from Takeda et al. (2010), which are the solar-analog stars. {\bf b} : All symbols are the same as {\bf a}, but the data points of superflare stars are limited to solar-analog (5,500K $<$ $T_{\rm eff}$ $<$ 6,000K, $\log g$ $\geq$ 4) superflare stars.}\label{fig:exbi}
\end{figure}

\begin{figure}[htbp]
 \begin{center}
  \FigureFile(120mm,120mm){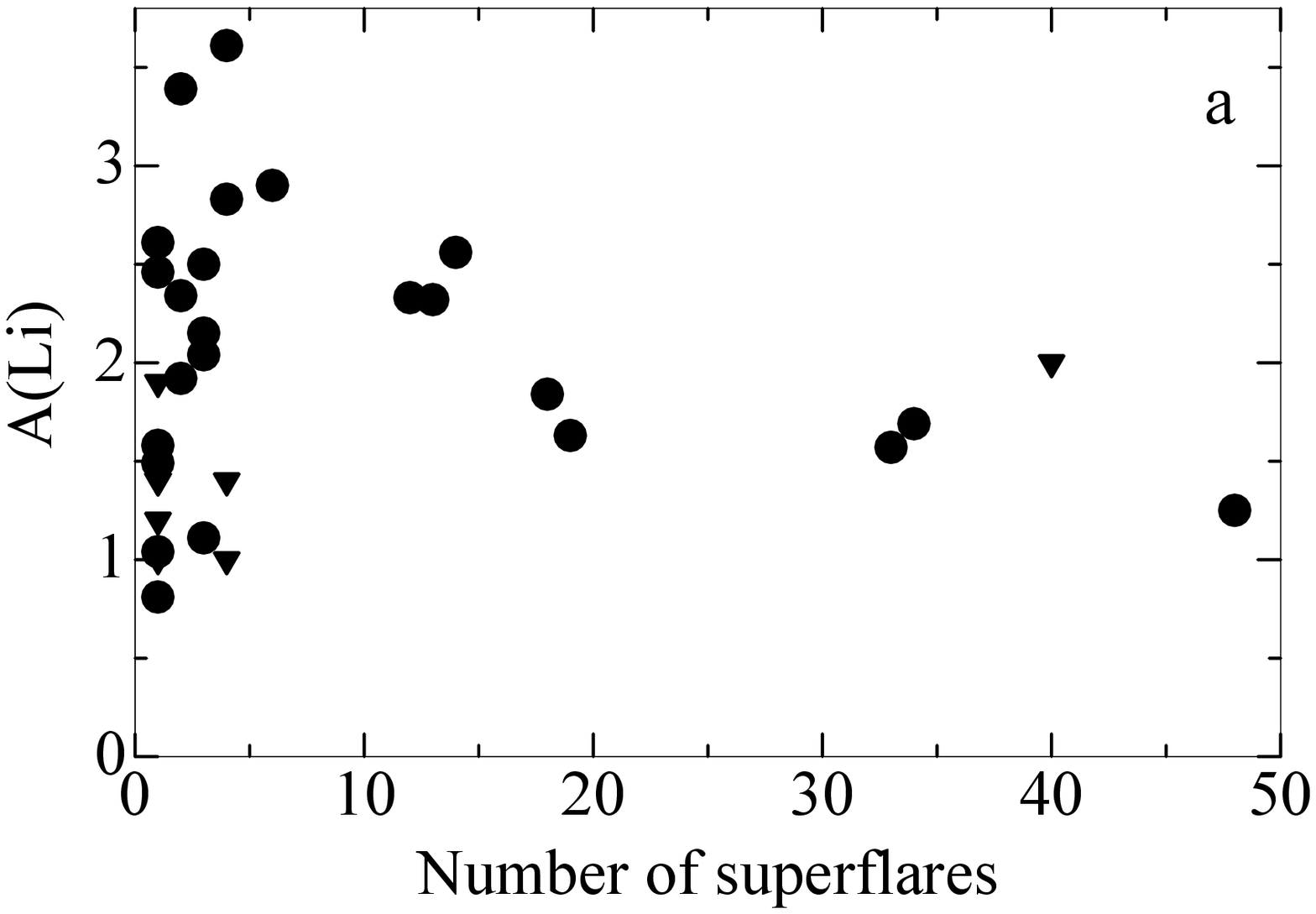} 
  \FigureFile(120mm,120mm){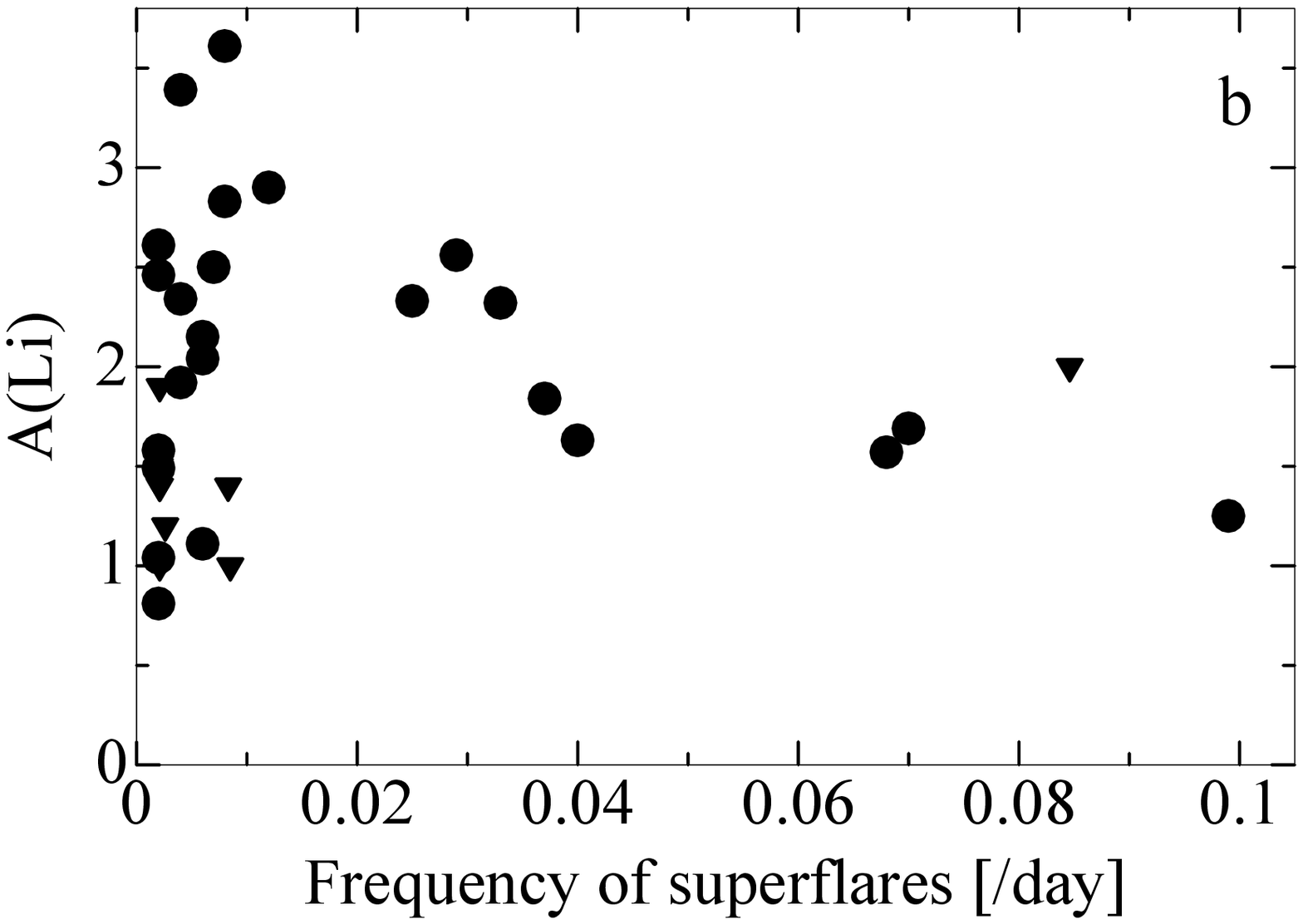} 
 \end{center}
\caption{{\bf a} : $A$(Li) vs. the number of superflares (Shibayama et al. 2013). The symbols of filled circles and triangles indicate the measurements of this work. The triangles indicate the upper limit of Li abundance of superflare stars.  {\bf b} : $A$(Li) vs. the frequency of superflares (the number of superflares divided by total observation time).}\label{fig:exbi}
\end{figure}

\clearpage

\begin{table}
	\begin{center}
	\caption{Stellar parameters, estimated rotation period, $v \sin i$, Li abundance, $r_{0}$ index, number of superflares, and total observation time.}\label{tab:Sppara}
	\begin{tabular}{rcccrcccccccl}
	\hline
	KIC ID & $T_{\rm eff}$\footnotemark[1] & $\log g$\footnotemark[1] & $v_{\rm{t}}$\footnotemark[1] &  [Fe/H]\footnotemark[1] & $A$(Li) &  $v \sin i$\footnotemark[1] & $P$\footnotemark[1] & $r_{0}$(8542)\footnotemark[2] & N\_f\footnotemark[3] & T\_obs\footnotemark[4] & Remarks \\
	& [K] & [cm s$^{-2}$] & [km s$^{-1}$] & & & [km s$^{-1}$] & [days] &  & & [days] & \\
	\hline
 3626094        &  6026   &  4.15   &  1.17   &  --0.03   &    2.56    &  2.9   &   0.7   &  0.24  &  6 & 483 &$i<20$\footnotemark[5]\\
 4742436	&  6033   &  4.20   &  1.09   &  --0.15   &    2.56    &  3.6   &   2.3   &  0.30  & 14 & 483 &$i<20$\footnotemark[5]\\
 4831454	&  5672   &  4.59   &  1.13   &    0.04   &    2.83    &  2.1   &   5.2   &  0.50  &  4 & 483 &$i<20$\\
 6503434	&  5902   &  3.63   &  0.96   &  --0.20   &    2.37    &  5.5   &   3.9   &  0.27  & 13 & 389 &\\
 6504503	&  5484   &  4.56   &  0.77   &    0.23   &    $<$1.2  &  1.6   &  31.8   &  0.25  &  1 & 389 &$i<20$\footnotemark[5]\\
 6865484	&  5800   &  4.52   &  1.11   &  --0.12   &    2.33    &  2.7   &  10.3   &  0.48  & 12 & 473 &\\
 6934317	&  5694\footnotemark[6]   &  4.42\footnotemark[6]   &  0.87\footnotemark[6]   &  --0.03\footnotemark[6]   &    1.25\footnotemark[6]    &  1.9\footnotemark[6]   &   2.5\footnotemark[6]   &  0.33\footnotemark[6]  & 48 & 483 &$i<20$\footnotemark[5] \\
 7093547	&  5497   &  4.62   &  0.18   &    0.26   &    1.04    &  2.1   &  14.2   &   -    &  1 &473 &\\
 7354508	&  5620   &  4.09   &  0.92   &  --0.10   &    0.81    &  3.2   &  16.8   &  0.40  &  1 &473 &\\
 7420545	&  5355   &  3.56   &  1.30   &  --0.03   &    1.57    &  6.3   &  36.2   &  0.40  & 33 &483 &\\
 8359398	&  5214   &  4.85   &  0.75   &  --0.13   &    1.11    &  1.8   &  12.7   &   -    &  3 &473 &\\
 8429280	&  5055\footnotemark[7]  &  4.41\footnotemark[7]  &  1.00\footnotemark[7]  &  --0.02\footnotemark[7]  &    2.9\footnotemark[7]    & 37.1\footnotemark[7]  &   1.2   &  0.68 &  6 &  483 &\\
 8547383	&  5606   &  4.38   &  0.83   &  --0.02   &    $<$2.0  &  4.4   &  14.8   &   -    & 40 &473 &\\
 8802340	&  5529   &  4.43   &  1.25   &  --0.13   &    1.92    &  4.3   &  10.3   &  0.49  &  2 &483 &\\
 9412514	&  6075   &  3.72   &  1.55   &    0.10   &    1.58    &  8.1   &   3.7   &  0.22  &  1 &483 &$i<20$\footnotemark[5]\\
 9459362	&  5458   &  3.82   &  1.39   &  --0.78   &    1.63    &  7.0   &  12.6   &   -    & 19 &473 &\\
 9583493	&  5605   &  4.34   &  1.28   &  --0.12   &    2.61    &  7.2   &   5.5   &  0.51  &  1 &483 &\\
 9652680	&  5618   &  4.80   &  1.00   &  --0.30   &    3.39    & 34.2   &   1.5   &  0.68  &  2 &483 &\\
 9766237	&  5606\footnotemark[8]   &  4.25\footnotemark[8]   &  0.88\footnotemark[8]   &  --0.16\footnotemark[8]   &    $<$1.4\footnotemark[8]  &  2.1\footnotemark[8]   &  14.2   &  0.26\footnotemark[8]  &  1 &473 &\\
 9944137	&  5666\footnotemark[8]   &  4.46\footnotemark[8]   &  0.93\footnotemark[8]   &  --0.10\footnotemark[8]   &    $<$1.4\footnotemark[8]  &  1.9\footnotemark[8]   &  12.6   &  0.27\footnotemark[8]  &  1 &473 &\\
 10252382	&  5334   &  4.01   &  1.40   &  --0.61   &    2.04    &  5.5   &  16.8   &   -    &  3 &473 &\\
 10387363	&  5381   &  4.75   &  0.65   &    0.07   &    $<$1.5  &  1.5   &  29.9   &   -    &  1 &473 &\\
 10471412	&  5776   &  4.53   &  0.91   &    0.15   &    $<$1.4  &  2.7   &  15.2   &  0.35  &  4 &483 &\\
 10528093	&  5180   &  3.96   &  0.71   &  --0.14   &    1.84    & 11.4   &  12.2   &  0.67  & 18 &483 &\\
 11140181	&  5552   &  4.60   &  1.25   &  --0.09   &    1.49    &  3.6   &  11.5   &  0.56  &  1 &483 &\\
 11197517	&  5284   &  4.59   &  0.62   &    0.42   &    $<$1.0  &  0.9   &  14.3   &  0.19  &  1 &483 &\\
 11303472	&  5193   &  4.57   &  0.93   &  --0.14   &    $<$1.0  &  2.1   &  13.5   &  0.49  &  4 &473 &\\
 11390058	&  5921   &  4.43   &  1.11   &  --0.15   &    2.50    &  3.5   &  12.0   &  0.32  &  3 &418 &\\
 11455711	&  5631   &  3.85   &  1.16   &  --0.23   &    2.15    &  7.9   &  13.9   &  0.40  &  3 &473 &\\
 11494048	&  5907   &  4.49   &  1.03   &    0.01   &    2.46    &  3.4   &  14.8   &  0.32  &  1 &408 &\\
 11610797	&  6209   &  4.41   &  1.70   &    0.26   &    3.61    & 22.5   &   1.6   &  0.63  &  4 &483 &\\
 11764567	&  5736   &  4.02   &  1.71   &  --0.08   &    1.69    & 14.7   &  22.4   &  0.49  & 34 &483 &\\
 11818740	&  5500   &  4.84   &  0.80   &    0.00   &    $<$1.9  &  3.0   &  15.4   &   -    &  1 &473 &\\
 12266582	&  5576   &  4.61   &  1.20   &  --0.06   &    2.34    &  4.8   &   6.9   &  0.52  &  1 &483 &\\
\hline
	\multicolumn{8}{l}{\hbox to 0pt{\parbox{170mm}{\footnotesize
    \footnotemark[1] The atmospheric parameters and brightness variation periods ($P$) which are estimated in Notsu et al. (2015a : Paper I).\\
	\footnotemark[2] Normalized intensity of the line center of Ca II 8542 line taken from Notsu et al. (2015b : Paper II).\\
	\footnotemark[3] N\_f is the number of superflares (Shibayama et al. 2013).\\
	\footnotemark[4] T\_obs is the observation time of each superflare stars by Kepler (Q0 - Q6).\\
	\footnotemark[5] Notsu et al. (2015b :Paper II).\\
	\footnotemark[6] Notsu et al. (2013b).\\
 	\footnotemark[7] Frasca et al. (2013).\\
	\footnotemark[8] Nogami et al. (2014).\\
}}}
\\
	\end{tabular}
	\end{center}
\end{table}

\clearpage

\begin{table}
	\begin{center}
	\caption{Stellar parameters, estimated rotation period, $v \sin i$, and Li abundance.}\label{tab:Sppara}
	\begin{tabular}{lccccrccc}
	\hline
ID & $T_{\rm eff}$\footnotemark[1] & $\log g$\footnotemark[1] & $v_{\rm{t}}$\footnotemark[1] &  [Fe/H]\footnotemark[1] & $v \sin i$\footnotemark[1] & $A$(Li) &\\
 & [K] & [cm s$^{-2}$] & [km s$^{-1}$] &  & [km s$^{-1}$] & &\\
\hline
 59 Vir     & 6127 & 4.32 & 1.38 &  0.19 &  6.2 & 2.92  \\
 61 Vir     & 5581 & 4.52 & 0.89 &  0.01 &  1.2 & $<$3.0  \\
 18 Sco     & 5760 & 4.37 & 1.02 &  0.03 &  2.3 & 1.50  \\
 HD163441	& 5768 & 4.45 & 0.87 &  0.05 &  2.5 & 1.55  \\
 HD173071	& 5982 & 4.41 & 0.98 &  0.18 &  2.7 & 2.48  \\
 HIP71813	& 5786 & 4.30 & 1.07 &  0.01 &  2.3 & 0.49  \\
 HIP76114	& 5746 & 4.50 & 0.98 &  0.02 &  1.8 & 0.76  \\
 HIP77718	& 5830 & 4.37 & 0.97 & -0.11 &  2.5 & 2.28  \\
 HIP78399	& 5830 & 4.44 & 1.04 &  0.07 &  2.5 & 1.83 \\
 HIP100963  & 5834 & 4.56 & 1.07 &  0.01 &  2.4 & 1.70  \\
 Sun(Moon)	& 5783 & 4.44 & 0.85 &  0.03 &  2.4 & $<1.0$,~0.92\footnotemark[2]  \\
\hline
	\multicolumn{8}{l}{\hbox to 0pt{\parbox{170mm}{\footnotesize
    \footnotemark[1] The stellar parameters are taken from Notsu et al. (2015a:Paper I).\\
    \footnotemark[2] The Li abundance of solar value are taken from Asplund et al. (2009).\\
}}}
\end{tabular}
\end{center}
\end{table}

\end{document}